\documentclass[12pt]{article}

\usepackage[utf8]{inputenc}
\usepackage{graphicx}
\usepackage{comment}
\usepackage{natbib}
\usepackage{amsmath}

\usepackage{lmodern}
\usepackage[top=4cm, bottom=4cm, left=3cm, right=2.5cm]{geometry}

\usepackage{setspace}
\usepackage{footmisc}

\makeatletter
\newcommand\iraggedright{%
\let\\\@centercr\@rightskip\@flushglue \rightskip\@rightskip
\leftskip\z@skip}
\makeatother
\iraggedright

\usepackage[T1]{fontenc}

\title{No Time for Time from No-Time}
\author{Eugene Y. S. Chua\thanks{eychua@ucsd.edu}
\and
Craig Callender\thanks{ccallender@ucsd.edu}}
\date{}

\begin{document}

\maketitle
\doublespacing
\section*{Abstract}
    \noindent Programs in quantum gravity often claim that time emerges from fundamentally timeless physics. In the semiclassical time program time arises only after approximations are taken. Here we ask what justifies taking these approximations and show that time seems to sneak in when answering this question. This raises the worry that the approach is either unjustified or circular in deriving time from no–time. 

\section*{Acknowledgements}

We thank Maaneli Derakhshani, Valia Allori, the Southern California Philosophy of Physics Group, and participants of the Workshop in Celebration of David Albert's Birthday for their comments/feedback. 

\newpage

\begin{abstract}
    \noindent Programs in quantum gravity often claim that time emerges from fundamentally timeless physics. In the semiclassical time program time arises only after approximations are taken. Here we ask what justifies taking these approximations and show that time seems to sneak in when answering this question. This raises the worry that the approach is either unjustified or circular in deriving time from no–time. 
\end{abstract}

\section{Introduction}
Programs in quantum gravity often produce supposedly fundamentally timeless formalisms. Because we observe change, it's important that they recover time from no-time somehow. One popular idea suggests time emerges from fundamentally timeless physics just as perceived color arises from the fundamentally uncolored world of basic physics. In canonical quantum gravity's semiclassical time program, the idea is that time emerges from fundamentally timeless physics after taking semiclassical approximations. Nothing fundamentally plays the "time role" throughout any solution, but time emerges in approximately classical sectors of some solutions.

Comparisons with perceived color suggest an obvious worry: circularity. Physically, color only emerges from uncolored matter diachronically. Color arises from observers like us interacting with matter across temporal intervals. Replace color with time and the threat is obvious: if time emerges from no-time but emergence requires time, then we can’t really say we’ve derived time from no-time. Time emerges if we blur our vision, but if blurring takes time then time never disappeared. 

Here, we raise this concern in a sharp way for the semiclassical time program. Focusing specifically on the approximations necessary to derive time from no-time, we'll show that time implicitly sneaks back in via the physical justifications behind these approximations. This leaves the program either unjustified in applying the approximations because we are applying them to timeless solutions, or succeeding only on pains of circularity. 

\section{The Problem of Time and Emergence of Semiclassical Time}

Quantum gravity seeks to reconcile our best theory of gravity, general relativity, with our best theory of matter, quantum theory. Different strategies exist, but we focus on the oldest canonical approach, quantum geometrodynamics, and its recovery of semiclassical time. We chose this program because it has been rigorously developed. We expect, however, that many lessons will generalise.

Canonical approaches employ a quantised Hamiltonian formalism. One therefore casts general relativity into its Hamiltonian “3+1” form, decomposing spacetime into leaves of spacelike hypersurfaces. The Hamiltonian framework demands canonical variables and conjugate momenta. For gravity, the basic variable is the three-dimensional spatial metric characterizing spacelike hypersurfaces. Its conjugate momentum is defined in terms of the trace of the spatial three-metric's extrinsic curvature. In classical mechanics the Hamiltonian governs the spatial configuration of particles through time; in classical Hamiltonian general relativity, the Hamiltonian governs the spatial geometry itself through time. Once put in this form, we quantise.

The counterpart of the quantum state is a functional operating in a configuration space of spatial three-metrics. To quantise, we turn the variables into operators. Trouble arises because general relativity is a constrained Hamiltonian system. One of the constraints is due to general relativity's time reparameterization freedom  -- we can foliate spacetime in many different ways. This constraint, the Hamiltonian constraint, demands that the Hamiltonian vanishes. Making the Hamiltonian an operator and imposing the constraint yields:

\begin{equation}
    \hat{H}\Psi(h_{ab}(x), \phi) = 0
\end{equation}

\noindent i.e., the famous Wheeler-DeWitt (\textbf{WD}) equation, where $\hat{H}$ is the Hamiltonian operator for both gravity and matter, and $\Psi$ is the \textbf{WD} wave-functional depending on the spatial three-geometries encoded by the spatial metric $h_{ab}(x)$ and whatever matter fields we include, e.g., $\phi$, a massive scalar field.

The semiclassical time program's core idea is that time emerges if $h_{ab}(x)$ is semiclassical. If not — if $h_{ab}(x)$ is quantum — then the concept of time won't find any realizer. This idea was expressed by DeWitt (1967) but developed by Banks (1985) in the canonical approach.\footnote{See Kiefer 2004, Anderson 2007 and references therein.} The \textbf{WD} wave-functional is, at the fundamental level, utterly timeless. Nonetheless it describes patterns of correlations, just like a checkered shirt at an instant contains a spatial pattern of correlations amongst stripes and colors. In the semiclassical interpretation, the idea is that at a certain level of approximation, a pattern of correlations ``looks” temporal, just as a checkered shirt can look solidly colored if one zooms out far enough.

By ``looks temporal” we mean that a parameter plays the time role. While defining the time role could become quite messy and philosophical, this program adopts a very minimal sufficient condition that seems plausible, namely, that something plays the time role if it behaves as ``$t$” does in the ordinary time-dependent Schrödinger equation (\textbf{TDSE}). In other words, if the matter fields vary with some parameter the same way they do with ``$t$” in the \textbf{TDSE}, that warrants calling that parameter time. 

Herein lies the key achievement of the semiclassical time program: given suitable approximations, they show that the non-temporal gravitational fields $h_{ab}$ can play the time role in a functional Schrödinger equation for the matter fields $\phi$. If one approximates from the \textbf{WD} equation appropriately, it looks like matter is evolving with respect to time (\textit{a la} Schrödinger equation) against a classical gravitational curved spacetime background (described by the semiclassical Einstein-Hamilton-Jacobi equation).

Let's turn to the actual derivation of time and the functional Schrödinger equation. Here we loosely follow a presentation by Derakhshani (2018). The derivation has two crucial steps. One, it uses the Born-Oppenheimer approximation (\textbf{BO}) to motivate factorizing the wave-functional of the universe. Two, it employs the \textbf{WKB} approximation on the gravity term in this product. One can think of the first move as separating out a sub-system from the total system. The second move shows that when that sub-system behaves approximately classically, it can function as a clock for the rest of the system.

Suppose we have a wave-functional that satisfies the \textbf{WD} equation and other necessary constraints. This describes a static wave in a high-dimensional configuration space. How do we get time?

To begin, notice that we don't expect quantum gravitational effects except near the Planck scale. Since $h_{ab}$ depends on the  extremely small Planck mass $m_p$, the idea of separating scales via (\textbf{BO}) is natural. Hence we can separate the ``heavy'' part of the wave-function, $\chi(h_{ab})$  from the ``light'' part, $\psi(\phi, h_{ab})$:

\begin{equation}
    \Psi\approx\chi(h_{ab})\psi(\phi, h_{ab})
\end{equation}

\noindent The idea is to use the $h_{ab}$ degrees of freedom as a clock for the light part $\phi$. 

We now apply a \textbf{WKB} approximation, substituting the ansatz $Ae^{iS}$ for a wave-function. We do that for the first factor, the heavy subsystem, turning the wavefunction into

\begin{equation}
   \Psi\approx A(h_{ab})e^{im_{p}^{2}S(h_{ab})}\psi(\phi, h_{ab})
\end{equation}

\noindent Next, expand S($h_{ab}$) as a power series in $m^2_p$:

\begin{equation}
S=m_{p}^{2}S_{0}+S_{1}+m_{p}^{-2}S_{2}... 
\end{equation}

\noindent Then, as usual in \textbf{WKB}, we plug $S_0$ and $S_1$ terms back into the wave equation and solve. In the ordinary quantum mechanical case, the $0^{th}$ order terms returns a Hamilton-Jacobi equation and the $1^{st}$ order term returns a continuity equation. Essentially the same happens here. Notably, solving to leading order $m_{p}^{2}$, we derive a semiclassical gravitational Hamilton-Jacobi equation.

Take a solution of these equations. Based on experience with geometric optics and quantum theory, we know it defines in superspace a vector field whose integral curves can be parametrized by a time. With this in mind, we define

\begin{equation}
    \dot{h}_{ab} = 2NG_{abcd}\frac{\delta S}{\delta h_{cd}} + D_a N_b + D_b N_a
\end{equation}

\noindent where $G_{abcd}$ is the DeWitt metric, $N$ is the lapse function, $D_a$ and $D_b$ are the spatial derivatives, and $N_a$ and $N_b$ are shift vectors. One now takes the matter wavefunction $\psi(\phi,t;h_{ab})$ and uses (5) to define a time derivative for it

\begin{equation}
    \frac{\partial \psi(\phi,t)}{\partial t} = \intop_{\Sigma} \dot{h}_{ab}(\textbf{x}, t)  \frac{\partial}{\partial h_{ab}}\psi(\phi, h_{ab})d^3 x 
\end{equation}

\noindent Time emerges in terms of this directional derivative. Call this \textbf{WKB time}.

The final step uses \textbf{WKB} again, keeps only the lowest order terms, and requires a lot of massaging. Skipping these details, we can show that $\psi(\phi,h_{ab})$ satisfies a functional Schrödinger equation

\begin{equation}
    i\frac{\partial}{\partial t}\psi(\phi,t;h_{ab})=\hat{H}^{m}(\phi;h_{ab})\psi(\phi,t;h_{ab})
\end{equation}

\noindent where $\hat{H}$  is a Hamiltonian-type term and $\psi$ is evaluated at a solution $h_{ab}$, which is itself a solution of the classical Einstein equations. 

Such a compressed derivation may be confusing to unfamiliar readers. The important take-away here is that the $t$ we used to parametrize the approximately classical general relativistic solutions (corresponding to the first ``heavy'' term in our factorization (2)) is used in a solution as a clock for the matter fields in the \textbf{WKB} regime. The “$t$” in (7) is the same as that in (6). We won't delve into the rest of the theory; however, note that we can also derive a continuity equation that allows us to use the normal Born rule for predictions from the theory, and furthermore, using perturbation theory -- by considering higher-order terms we have so far ignored -- one can derive non-classical predictions. 

In sum, the semiclassical derivation provides an elegant derivation of time from no-time. Making a series of seemingly reasonable assumptions, a parameter that looks and acts like time emerges. And if we agree that something that looks and acts like time \textit{is} time, then time emerges.

\section{Justifying the Approximations}

We jumped from one equation to another by expanding to leading order, focusing on lowest order, assuming the wave-functional approximately factorises, and so on. What justifies these steps? Approximations require physical justification. At the level of pure math, one can “derive” virtually any equation from any other if allowed to assume anything. It makes no sense to say that one equation or quantity is ``close'' to another absent a metric. We need justification, and it is in this physical justification that we fear time sneaks in. 

To elaborate,  we can treat classical pendulums as approximately undamped harmonic oscillators. For small angles, $sin\textbf{ }\theta \approx \theta$, allowing us to derive equations of motion for pendulums which are identical to that of harmonic oscillators. A harmonic oscillator, we might say, ``emerges'' from the pendulum in the small angle limit. But relative to some measurement standard, at some point an initial displacement angle becomes too big and the approximation fails; that is, we notice deviations from the derived equation of motion. Angles aren't intrinsically big or small. They are big or small relative to a standard. Typically that standard refers to the observational or measurement capacities of an observer. The approximation's validity hangs partly on an error analysis of our measurement technique. Coarse measurements allow the approximation to be good for greater values of $\theta$ than finer measurements. 

This example suggests a subtle problem for the semiclassical time program and even the present analysis. We have no observers yet in canonical quantum gravity. We are working in a partially interpreted theory, one lacking a solution to the infamous measurement problem. Absent observers, we cannot perform the above error analysis. When are (say) off-diagonal terms in matrices ``small'' and justifiably ignored? The answer: when they're irrelevant to the observer (measurement/analysis/etc.). However, to introduce an observer in order to have a standard for judging smallness, we effectively already need time. Observation is a temporal process. So we can only justify approximations by already introducing time, making the derivation circular. Sans an observer, we can't say what ``looks'' like a small difference that would warrant an approximation. 

We'll return to this point, but for now we keep things simple by noting that the approximations used to derive semiclassical time are always warranted in the rest of physics by appeal to an implicit time metric. Without the time metric, the approximations seem physically unwarranted. We do not, and cannot, show that there is \textit{no standard possible} warranting these approximations. What we can do is raise the worry and challenge advocates of semiclassical time to justify the approximations without appealing to a prior time standard. We'll see that some apparently innocent assumptions are, in fact, not.

Although the semiclassical time program has an estimated twenty assumptions (Anderson 2007), we'll concentrate on three: the Born-Oppenheimer approximation, the \textbf{WKB} approximation, and decoherence.

\subsection{The Born-Oppenheimer Approximation} 

The \textbf{BO} approximation in the semiclassical time program splits the universe into two kinds of subsystems, the gravitational field $h$ and quantum matter fields $\phi$. The justification for this split ultimately appeals to a difference in masses: masses associated with $h$ are `heavy' in comparison to the masses associated with $\phi$.\footnote{See e.g. Banks (1985, 337--338), Kiefer (2004, 165).} Therefore it seems plausible that $h$ is largely insensitive to $\phi$. By contrast, $\phi$, being small and light, is sensitive to the big and heavy $h$. We therefore assume that the wave-functional $\Psi$ for the entire system (the universe) can be approximately factorized into two wave-functions $\chi(h)$ and $\psi(\phi, h)$, with $\chi$ associated with the heavier $h$, and $\psi$ associated with both the lighter $\phi$ \textit{and} $h$, as per (3). This factorization, as we've seen, is a necessary assumption in the above derivations. 

On its face the rationale doesn't sneak time in. Some masses are larger than others, and we expand accordingly. That's it.

Let's probe deeper. \textbf{BO} is motivated by appealing to the ``very different scales'' (Kiefer 2004, 164) that the gravitational fields and matter fields have. This appeals to a metric that measures how big the effects of one subsystem are on the other. Why does having different size masses warrant different scales and factorizing the wavefunction? Differences in the values of other properties (say, charge) don't always demand or legitimize such an approximation. What is special about mass?

To help answer this question let's look at standard uses of \textbf{BO} outside quantum gravity. Unfortunately we'll find that mass and size scale differences between systems are only relevant for \textbf{BO} because they are proxies for \textit{timescale differences} in the dynamics of the relevant subsystems. 

In its most popular application -- molecular and atomic physics -- \textbf{BO} is used to factorize an atom or molecule's wave-function into the product of two subsystems. Here, the heavier subsystem is the nuclei, and the lighter subsystem is the electrons surrounding the nuclei (Griffiths 2005). Again, the heavier system is assumed to be effectively independent of the lighter system, while the lighter system rapidly adapts itself to changes in the heavier system. Usually, we pretend that the nuclear wave-function is not changing at all \textit{in time}, and then calculate the electronic wave-function associated with that nuclear wave-function. We then find a more realistic nuclear wave-function by letting it vary `slowly' or `sluggishly', calculating the possible ranges of electronic wave-functions and hence the mean potentials in which the nuclei can move.

More generally, \textbf{BO} applies in cases where heavier subsystems are known to change slowly \textit{in time} with respect to lighter subsystems. That is why mass matters. Heavier subsystems have significantly different \textit{characteristic dynamical timescales} -- timescales over which ``the parameters of the system change appreciably'' -- with respect to lighter subsystems, and can be said to be \textit{adiabatic} with respect to the lighter subsystem. The change in the lighter subsystem happens on such a short timescale that there isn't enough time for the heavier subsystem to react in that relevant timescale, and so it is effectively independent of lighter subsystems in that period of time. \textbf{BO} is thoroughly laden with temporal notions.

Returning to the semiclassical time program, a problem arises. Because \textbf{BO} is so widely used, and because it initially seems to be about mass (not time!), it may be imported into derivations without considering whether its use in new applications is warranted. Did that happen here? We cannot say, but we leave this section with a dilemma: either the mass scales relevant here are proxies for time scales or not. If they are, we face circularity; if they are not, we have no clear means of assessing whether \textbf{BO} is even applicable here. In short, this seems to be a case of needing time to get time, but of course, in canonical quantum gravity we have no time for that.

\subsection{The \textbf{WKB} Approximation}

The \textbf{WKB} approximation is a staple of every quantum mechanics course. Often presented as a piece of pure math, \textbf{WKB} seems like a mere approximation method in the theory of partial differential equations, an unlikely place to find a hidden time preference. But of course, we still need physical justifications for why this math applies to a given physical situation. For that we need physics.

Frequently, we use \textbf{WKB} when working with stationary states of energy $E > V$. Immediately, we note that the time dependence is therefore hidden. If a system begins in an energy eigenstate, then time evolution simply multiplies the state by a time-dependent phase factor that doesn't affect the probabilities for measurement. Perhaps we shouldn't think this way: here, the time-independent equation is fundamental and the time-dependent one is non-fundamental, contrary to ordinary quantum theory.

Still, we believe the approximation presumes the existence of time. We see this most clearly with the textbook \textbf{WKB} derivation. Begin with the one-dimensional time-independent Schr$\"o$dinger equation (\textbf{TISE}) describing a system in a background potential $V(x)$:

\begin{equation}
\frac{d^{2}\psi}{dx^{2}}+\frac{2m}{\hbar^{2}}(E-V(x))\psi=0
\end{equation}

\noindent or:

\begin{equation}
\frac{d^{2}\psi}{dx^{2}} + \frac{p(x)^2}{\hbar^2}\psi = 0
\end{equation}

\noindent where we use the classical momentum identity: 

\begin{equation}
p(x) = \sqrt{2m(E - V(x))}
\end{equation}

\noindent If $V(x)$ is constant, the system behaves like a free particle with $\psi(x) \sim e^{ip(x)}$. If $V(x)$ varies \textit{slowly}, we expect that the system behaves \textit{approximately} like a free particle. Motivated by this, we find solutions to the \textbf{TISE} such that

\begin{equation}
\psi(x)=A(x)e^{iS(x)/\hbar}
\end{equation}

\noindent Plugging this back into the \textbf{TISE}, we get two equations (for the imaginary and real parts respectively):

\begin{equation}
\hbar\frac{d^{2}A}{dx^{2}} =  A\left((\frac{dS}{dx})^2 - \frac{p(x)^2}{\hbar^2}\right) 
\end{equation}

\begin{equation}
2 \frac{dA}{dx}\frac{dS}{dx} + A\frac{d^{2}S}{dx^2} = 0
\end{equation}

\noindent Everything so far is exact. However, note that (12) \textit{generally does not have analytic solutions}. What then? The solution, and a crucial step in \textbf{WKB}, is to \textit{assume} that $A$ varies \textit{so slowly} with respect to $x$ that $\frac{d^{2}A}{dx^{2}} \approx 0$.

This step allows us to solve (12) and (13) for $A$ and $S$. Combining these results, we get the well-known \textbf{WKB} approximation to the wave-function

\begin{equation}
\psi(x) \approx \frac{C}{\sqrt{p(x)}}exp\left(\pm\frac{i}{\hbar}\intop dx\textbf{ }p(x)\right)
\end{equation}

\noindent where $C$ is some real constant dependent on $A$ and $S$. Arbitrary superpositions of these wave-functions are approximate solutions of the Schrödinger equation. They are also exact solutions of the classical Hamilton-Jacobi equation — from which one obtains the time parameter used in the semiclassical time program. 

Under what conditions are we allowed to neglect $\frac{d^{2}A}{dx^{2}}$? This is where the physics enters. The answer is well-known: $V$ must vary slowly with $x$ and ($E - V$) can't be  too small. When $V$ is constant, and the system behaves like a free particle, $A$ is constant. When $V$ is `close to constant', i.e. varying slowly, so too is $A$.

On its face the condition of $V$ “slowly varying” does not conceal any time-dependence since it concerns slowness with respect to the spatial $x$ not the temporal $t$. What motivates \textbf{WKB} is that when the potential is not too spatially sharp one tends to not see much interference, so this important assumption is about spatial smoothness not temporal variation.

Still, time is present. There are many ways to see this. An obvious one is to consider the use of the classical momentum identity (10). In quantum mechanics, we know that the momentum operator depends only on spatial variables and not time: 

\begin{equation}
    \hat{p} = -ih\nabla
\end{equation}

\noindent However, the classical momentum \textit{does} depend on time, since: 

\begin{equation}
    p = m\frac{dx}{dt}
\end{equation}

\noindent Despite working in quantum mechanics, we used the classical momentum identity without explanation. This lets us adopt the energy condition, $E > V$ not being too small (and $ E \neq V$), to physically justify neglecting $\frac{d^{2}A}{dx^{2}}$. But \textit{why} are we considering $ E > V$? The time dependence of (10) lets us see why. Combining (10) and (16) and separating variables yields:

\begin{equation}
    \intop dt = \intop dx \frac{m}{\sqrt{2m(E - V(x))}}
\end{equation}

\noindent Now we see \textit{why} $E > V$ is the relevant condition for \textbf{WKB}. For any fixed potential $V(x)$, the integral on the right hand side is small when $E - V(x)$ is large. As a result, the total time $\Delta t = \intop dt$ spent by a system in that constant potential is very small. On the contrary, if $E - V(x)$ is not too large (but $E \neq V$), then the total time spent under a fixed $V(x)$ is relatively longer. The \textit{longer} a particle generally spends \textit{time} moving in each given fixed potential, the \textit{slower} we can say the potential is varying spatially. The latter fact lets us derive \textbf{WKB} - but notice how the temporal metric is involved in the physical justification.

One might worry that this imports `classical bias' about particles into quantum mechanics, but we see essentially the same point from a wave perspective. Note that if the potential spatially varies slowly with respect to the particle's de Broglie wavelength, then its wave-function approximates that of a free particle, i.e., a plane wave. That means the system will propagate freely with a constant velocity $v$ for a time $T$. As Allori and Zanghi (2009, 24) note, that time -- the time for which we can pretend that $V$ is effectively constant -- satisfies the following relation: 

\begin{equation}
    T \sim \frac{L}{v}
\end{equation} 

\noindent where $L$ is the scale of variation of the potential. This provides a clear physical picture of what it means to apply \textbf{WKB}. If $L$ is long and $v$ is low, then the particle is moving slowly through an effectively unchanging $V$, allowing \textbf{WKB} to hold for long times. Conversely, if $L$ is short and $v$ high, then the particle rapidly moves (in time!) through the potential -- in these cases we can no longer assume that $V$ is effectively constant for the system, and \textbf{WKB} will not hold for long times. This clearly parallels the classical case discussed earlier. 

Since $\lambda = \frac{\hbar}{p} = \frac{\hbar}{mv}$, we can write (18) as:

\begin{equation}
    T \sim \frac{Lm\lambda}{\hbar} 
\end{equation} 

\noindent The time-dependence, evident when talking about velocities/momenta, becomes masked when we replace velocities with notions of wavelengths and spatial variations. Yet the time-dependence is plainly there. From (18) and (19)  we can see that if $L$ is large, the \textbf{WKB} approximation will be good for long $T$ and if small then only for short $T$. 

In standard cases \textbf{WKB} is thus justified via a background time metric. In the case of semiclassical time, however, there is no such background time metric, so we again face our challenge to justify the assumption without invoking time.

\subsection{Decoherence}

The discerning reader might have noticed two sleights of hand in deriving the functional \textbf{TDSE} and `t'. First, in using \textbf{BO}, we effectively assumed that $\Psi$ was an eigenstate (3) of the \textbf{WD} equation. Since the \textbf{WD} equation is linear, general solutions involve a superposition of states. Second, a similar assumption was made in choosing the approximate \textbf{WKB} wave-function for the gravitational fields $\chi(h_{ab})$ in (3). Again, due to linearity, arbitrary superpositions of states are also solutions. These assumptions are absolutely vital for deriving a functional \textbf{TDSE}.\footnote{See Kuchar 1992.} Using an arbitrary superposition of states in the \textbf{BO} and \textbf{WKB} approximations, the above procedures do \textit{not} recover a semiclassical time.  

The most popular response to these observations appeals to \textit{decoherence}. (Kiefer 2004, 317) The idea is that if the initial state of the universe is in an arbitrary superposition of states, then decoherence will drive the wavefunction into a superposition of effectively non-interacting components, each one of which is suitable for the semiclassical time recovery. In an Everett-type interpretation of quantum mechanics, for instance, we could recover a time in each decohered branch or world. 

Our worry is especially clear here because decoherence is normally understood as a dynamical process. It presumes temporal evolution by the Schrödinger equation. Decoherence at once requires time and is required for time. Indeed, one finds tension in Kiefer's own account. On the one hand, he writes that ``A prerequisite [of decoherence in the semiclassical time program] is the validity of the semiclassical approximation... This brings an approximate time parameter t into play.'' (Kiefer 2004, 311) But later he writes that ``Since [decoherence] is a prerequisite for the derivation of the Schrödinger equation, one might even say that time (the WKB time parameter in the Schrödinger equation) arises from symmetry breaking [i.e decoherence]... Strictly speaking, the very concept of time makes sense only after decoherence has occurred.'' (Kiefer 2004, 318) Obviously, the two claims cannot be true at once, and again, we face our dilemma. 

\section{Discussion}

Our investigations into three approximations integral to the semiclassical time approach have unearthed a general worry: we seemingly need to put time \textit{in}, somewhere and somehow, in order to get time \textit{out} of the timeless formalism. This worry hasn't been noticed before, we suspect, because time is not blatantly assumed in derivations. It appears implicitly via the justifications for the assumptions, not explicitly in the math. 

Note that we haven't shown the impossibility of answering our challenge. If we could make sense of an atemporal observer, perhaps we could find a measurement standard that makes the terms ignored in \textbf{BO}, decoherence and \textbf{WKB} small in some relevant sense. Absent such observers, we observe that there is very little to work with in canonical quantum gravity to help us. This point becomes clearer by comparing our objection to a similar one leveled against decision-theoretic attempts to derive Born's rule in Everettian quantum mechanics. 

As is well-known, the Everettian interpretation faces a problem in making sense of quantum mechanical probabilities. Its law consists only of a linear deterministic wave equation. Therefore it produces only trivial probabilities (0, 1) for any outcome. Born's Rule, our guide to experiment, seems unexplained. In response, some Everettians have turned to decision theory, by trying to prove that rational Everettian agents will set their preferences in accordance with Born's Rule. Controversy ensues about whether the assumptions used in the proofs are really requirements of rationality. 

But another line of criticism will immediately sound familiar. Baker (2007), Kent (2010) and Zurek (2005) point out that Everettians use decoherence to say that different ``worlds'' approximately emerge from the wave-function. What does ``approximately'' mean here? Well, it seems to mean that a branching structure is likely to happen -- the probability of an error is small according to the Born measure (mod-squared amplitude). Yet the decision theoretic proofs begin with a branching structure. That begs the question, the critics say, for we've assumed that mod-squared amplitude is a probability in our demonstration that mod-squared amplitude is probability. 

Structurally this objection is similar to ours. Can any replies there be transferred to our case? 

The only Everettian response we found is Wallace (2012, 253--4). Wallace argues that the branching structure ``really is robustly present'' even prior to interpreting mod-squared amplitude as probability. What standard makes it present? His answer: Hilbert space norm. This is an objective physical measure. If branching emerges approximately with respect to Hilbert norm, then the probability measure is not needed as an assumption in deriving Born's rule. One could justifiably ask whether Hilbert space norm is enough to answer the objection. Small differences in Hilbert space norm may not be small differences for an observer, or vice versa. From color science we know that similar-looking colors (with small phenomenological distance) might be produced by physically dissimilar properties. Hilbert space norm might not be enough to fully answer the charge. 

However that debate goes, we lack anything like Hilbert space norm in the present case. The space of spatial three-metrics has a geometry given by the DeWitt metric. But this metric won't say how far quantum states are from one another. What we need, comparable to the Hilbert norm, is an invariant positive-definite inner product on \textbf{WD}'s solution space. Here we're right back to time! ``Invariant'' means the inner product is independent of time. Constructing an invariant positive-definite inner product on \textbf{WD}'s solution space is the notorious ``Hilbert space problem'' (Kucha$\check{r}$ 1992). While the Schrödinger equation provides a conserved inner product ``for free'', \textbf{WD} doesn't. The most natural way to solve the Hilbert space problem is to identify a time variable and construct a norm from that; but in this context that won't help. 

Again, we don't want to say that there is no way to warrant the approximations. But we have argued that the most natural warrant appears temporal. We see no reason to think the introduction of observers will change that verdict. 

\section{Conclusion}

We started with the idea that the world was fundamentally timeless: semiclassical time arises from certain regimes looking temporal when we blur our vision. That metaphor turns out to be not quite right, as it neglects that we've imported a mathematical construct, the Hamilton-Jacobi structure, onto the basic physics. Only within that structure does time seemingly emerge. Instead of blurry vision making a pattern of correlations in the wave-functions look temporal, what's happened is that we're being offered ``time glasses.'' We are told you're justified in using these glasses -- this mathematical construct -- and when we look through them, they turn the pattern temporal. Are we justified in wearing ``time glasses''? It seems the only reason to wear them is when one already has time. 

\bibliography{citation.bib}
\nocite{*}
\end{document}